# Designing the new architecture of international financial system in era of great changes by globalization

V.O. Ledenyov and D.O. Ledenyov

*Abstract* – We present a broad agenda for meaningful banking regulation reform aiming the creation of evolutive competitive environment to maximize the effectiveness of international financial system through the introduction of fair competition process among the banks in free market capitalism. We assume that the international financial system may evolve or decline within the evolutive competitive environment depending on both the environmental regulation policies as well as the competition between the banks. We present the commonly known definition of competition and apply the conceptual collateral thinking to identify the source of competitive strengths of financial institutions in free market capitalism. We explore how the banks conduct a search for competitive strategies. We show that the root cause of crisis in finances is hidden in the very wrong regulation policies and ideas behind these policies, which failed to create the evolutive competitive environment for effective, profitable, responsible and sustainable bank operation within existing international financial system. We review a number of initiatives on meaningful banking regulation reform proposed by central bankers from the G20 nations. We propose to introduce the Random Tax and the Quantum Tax. We argue that the introduction of the Random Tax and the Quantum Tax may compensate for the negative effects commonly associated with the existing banking regulation limitations imposed on the international financial system. We believe that the Random Tax and the Quantum Tax will improve the evolutive competitive environment and make it possible for the management teams at financial institutions to search for and to execute the winning virtuous business strategies toward the effective, profitable, responsible and sustainable banks operation.

The British multinational banking was created in the course of evolutionary processes in finances over 18, 19, and 20 centuries in Jones G G (1993). The competition among banks was a main driving force behind the evolution of British multinational banking. The powerful forces of competition shaped the competitive landscape in banking at national and global scales creating the traditional architecture of international financial system in Samuelson, Barnett (2007).

**So, let us provide a definition of competition** from The American Heritage Dictionary (1985):

1. Competition is the act of competing for a profit or a prize;

2. Competition is the rivalry between two or more businesses striving for the same customer or market.

Considering the modern international financial system, one question may arise in analogy with research approach in Chandler (2005): What is the base for the competitive strengths of financial institutions in free market capitalism?

In free market economies, the competitive strengths of financial institutions and investment firms rest on learned organizational capabilities in Chandler (2005). The process of organizational learning in financial institutions begins with the building of profit making commercial or investment bank, which is accomplished through the creation of organizational capabilities based on three types of knowledge in Chandler (2005):

1. Technical capabilities (knowledge of advanced technologies to create new financial products or services)

2. Functional capabilities (development, marketing and distribution);

3. Managerial capabilities (management knowledge and experience).

Evolving paths of learning in the financial industry are similar to other industries. They include the three stages in Chandler (2005):

1. Building Barriers to Entry;

2. Defining Strategic Boundaries of the competing banks within the financial industry competitive arena;

3. Encountering Limits to Growth.

The main problem to understand is: How do the commercial and investment banks create the barriers to entry and define the strategic boundaries to compete with other banks? The answer is that they do so by adopting a **corporate strategy** in Chandler (2005).

Focusing on the bank's strategy, we have to start with the two questions formulated in Gavetti, Rivkin (2007):

1. Where does a bank's competitive strategy come from?

2. How do the initial conditions, foresight, experience, competitive feedback, and other forces combine to shape the origins of competitive strategies?

**Let us define the strategy**: The strategy is a management team's way of seeing its place in its environment as well as the firm's way of interacting with the environment in Gavetti G, Rivkin J W (2007). Creation of strategy is cognitive time dependant process in Johnson, Scholes, Whittington (2007); Johnson, Langley, Melin, Whittington (2007); Johnson, Scholes, Ambrosini (1998); Johnson, Hendry, Balogun (1993). Usually, the management team conducts the strategy search using the positioning model in Porter (1980, 1985, 1996, 2008); Ghemawat (1991); Brandenburger and Stuart (1996) and the evolutionary model in Nelson, Winter (1982); Winter (1987, 2000). Coupling of the inductive logic in Wikipedia (2009), deductive logic in (Wikipedia) and abductive logic in Peirce (1958); Wikipedia (2009) is effectively applied in the positioning model and evolutionary model with the purpose of competitive strategy search by management team in organization in process of competition among banks in evolutive competitive environment in free market capitalism in Gavetti G, Rivkin J W (2007); Collis, Montgomery, Goold, Campbell, Prahalad, Lieberthal, Hart (1999).

**Let us provide the definitions for the different logics employed in strategy search process**:

Inductive logic — the logic of what is operative — reasons from the specific to the general in Martin (2008, 2009). Induction allows inferring $a$ entails $b$ from multiple instantiations of $a$ and $b$ at the same time in Wikipedia (2009).

Deductive logic — the logic of what must be — reasons from the general to the specific in Martin (2008, 2009). Deduction allows deriving $b$ as a consequence of $a$. In other words, deduction is the process of deriving the consequences of what is assumed in Wikipedia (2009).

Abductive logic –the logic of what could possibly be true –reasons through successive approximation in Martin (2008, 2009). Abduction allows inferring $a$ as an explanation of $b$, because of this, abduction allows the precondition $a$ to be inferred from the consequence $b$ in Wikipedia (2009).

**Let us make the definition of the competitive strategy execution / implementation**: The execution of competitive strategy from the organizational perspective is a chain of changes management by management team in a bank, which competes with other financial institutions in evolutive competitive environment in free market capitalism.

Now, we have completed our short review on:

1. Origin of competitive strategy;

2. Methods of search for competitive strategy; and learned about

3. True meaning of strategy execution definition.

Therefore, we can make an important conclusion that **the evolutive competitive environment has a direct impact on the ability of management team at bank to conduct the strategy search and make a choice for the best competitive strategy toward the effective, profitable, responsible and sustainable bank operation. In our opinion, most of the resources to create a basis for effective corporate strategy to gain competitive advantage over competing banks, are mainly derived by bank's management team from the evolutive competitive environment.**

Next practical question is: Can the banks be profitable by pursuing the best competitive strategies in conditions, when the evolutive competitive environment starts to degrade, because of imperfect regulation? The answer is "No." The existing banking regulation creates the evolutive competitive environment with its own limitations, and in turn, impacts the full process of strategy search and execution by management teams in financial institutions. Therefore, we can conclude that **the root cause of crisis in finances is hidden in the very wrong regulation policies and ideas behind these policies, which failed to create the evolutive competitive environment for effective, profitable, responsible and sustainable bank operation within existing international financial system, but not the inability of management teams of banks to find the competitive strategies.**

Let us think about the next question: How will the global financial crisis change the competition among the financial institutions?

We have to acknowledge the fact that the competitive landscape in banking was already changed completely as a result of crisis. Some big investment firms were bankrupt in the USA, the UK and elsewhere. Many banks turned for financial support from the central banks, which are the lenders of last resource.

We think that it is necessary to break up the big inefficient banks, because the small and medium size banks tend to be more efficient in serving the needs of small and medium size businesses. The small bank management teams do enormously benefit from flexible managerial structure during the decision making process and can react on constantly changing local market conditions promptly. In addition, the capital accumulation, supply and distribution chains are more optimized for purposes of commercial and investment banking in small banks. Therefore, we firmly believe that small and medium size banks will dominate and define the competitive landscape in banking in era of great changes by globalization.

Let us focus on the key problem: How will the proposed regulatory changes affect competition between banks and between banks and other institutions?

Fist of all, in our opinion, the meaningful banking regulation reform has to create a new evolutive competitive environment to maximize the effectiveness of international financial system through the introduction of fair competition process between the banks in evolutive competitive environment in free market capitalism. Speaking figuratively, we think that the management teams at banks could indeed be able to adapt and implement the virtuous competitive business strategies, if the genuinely evolutive high competitive environment for banks operation within international financial system would be created in first place.

The key issue is: How can the goal of creation of new regulation for evolutive competitive environment be achieved in conditions of globalization?

Let us take a look on the history of dealing with similar financial and economic downturns in the USA in the periods of great depression and recessions in last century. The Roosevelt administration engineered sweeping federal intervention into the marketplace in time of financial crisis, including in Thompson (2009):

1. Creation of federal deposit insurance;
2. Securities regulation;
3. Banking supervision; and
4. Separation of commercial and investment banking under the Glass-Steagall Act.

Let us review a number of initiatives on meaningful banking regulation reform proposed by central bankers from G20 nations at present time.

Ben S Bernanke, Chairman of the Board of Governors of the US Federal Reserve System proposes a broad-based agenda for reform, which should include at least five key elements in Bernanke (2009):

1. Consolidated supervision of systemically important financial institutions: Legislative change is needed to ensure that systemically important financial firms are subject to effective consolidated supervision, whether or not the firm owns a bank;

2. Systemic risk oversight: Oversight council made up of the agencies involved in financial supervision and regulation should be established, with a mandate to monitor and identify emerging risks to financial stability across the entire financial system, to identify regulatory gaps, and to coordinate the agencies' responses to potential systemic risks. To further encourage a more comprehensive and holistic approach to financial oversight, all federal financial supervisors and regulators – not just the Federal Reserve – should be directed and empowered to take account of risks to the broader financial system as part of their normal oversight responsibilities;

3. Improved resolution process: New special resolution process should be created that would allow the government to wind down a failing systemically important financial institution whose disorderly collapse would pose substantial risks to the financial system and the broader economy. Importantly, this regime should allow the government to impose losses on shareholders and creditors of the firm;

4. Payment, clearing, and settlement arrangements: All systemically important payment, clearing, and settlement arrangements should be subject to consistent and robust oversight and prudential standards;

5. Consumer protection: Policymakers should ensure that consumers are protected from unfair and deceptive practices in their financial dealings.

Also, Ben S Bernanke notes that the assets on the Federal Reserve's balance sheet have changed, because of financial crisis, and can be usefully grouped into four new categories in Bernanke (2009):

1. Short-term lending programs that provide backstop liquidity to financial institutions such as banks, broker-dealers, and money market mutual funds;

2. Targeted lending programs, which include loans to nonfinancial borrowers and are intended to address dysfunctions in key credit markets;

3. Holdings of marketable securities, including Treasury notes and bonds, the debt of government-sponsored enterprises (GSEs) (agency debt), and agency-guaranteed mortgage-backed securities (MBS); and

4. Emergency lending intended to avert the disorderly collapse of systemically critical financial institutions. I will say a bit more about each of these in turn.

Mario Draghi, Governor of the Bank of Italy and Chairman of the Financial Stability Board states in Draghi (2009): "Countries have responded to the severity of the crisis with discretionary policies to sustain the economy. Two main objectives for economic policy are:

1. Ensuring the resumption of progress towards consolidating the public finances, in order to converge towards budgetary balance; and

2. Sustaining productivity and economic growth.

Also, Mario Draghi stresses that the response to the crisis has consisted primarily in three decree laws, which together appropriated €25 billion of resources for the three years 2009-11 in Draghi (2009):

1. First package introduced transfer payments to low-income households (known as the family bonus, which supplemented the social card scheme launched during the previous summer); automatic income stabilizers were strengthened for the two years 2009-10 (a measure funded with national and EU funds on the basis of a State-Regions agreement); there were increases in the allocations for investment in public works and in investment grants to the State Railways group (in part to offset previous cutbacks), in conjunction with the adoption of new procedures to speed up the realization of projects included within the National Strategic Reference Framework.

2. Second measure issued last February dealt mainly with incentives for the purchase of cars;

3. Third decree law introduced tax incentives for investment in machinery between July 2009 and June 2010, through the exclusion from firms' income of half the expenditure.

The structural reforms are needed to render the measures efficacy:

1. Reform of the budget cycle may improve the control of expenditure;

2. Fiscal federalism must contribute to curbing expenditure;

3. Significant increase in the average effective retirement age is desirable;

4. Reduce tax evasion;

5. Reforms for reviving growth;

6. Increase the efficiency of government departments;

7. Improve the quality and effectiveness of government action;

8. Modernize schools and universities;

9. Eliminate the fragmentation of worker protection."

Axel A Weber, President of the Deutsche Bundesbank expresses his opinion in Weber (2009):

The financial crisis has demonstrated that banks remain the primary link between the financial system and the real economy.

While it is true that regulation and oversight have to be extended to all systemically important financial institutions, better regulation of banks is therefore an even more important building block in a stable financial system. For this reason, it is essential to further improve banks' risk management and to review their capital and liquidity requirements.

Svein Gjedrem, Governor of Norges Bank (Central Bank of Norway) makes his point on the issues in Gjedrem (2009):

1. Higher capital adequacy requirement and new minimum equity capital ratio requirement: Banks are required to hold capital in proportion to their risk-taking, the so-called minimum Tier 1 capital requirement. Tier 1 capital is composed of equity capital and other capital that can be used to absorb losses on a going-concern basis;

2. Strengthen quality of Tier 1 capital: Under the current rules, up to 50 per cent of a bank's Tier 1 capital can consist of hybrid instruments, which are a combination of debt and equity. The remainder must comprise common equity only. Perpetual capital securities and other hybrid instruments cannot be used as easily as equity capital to absorb losses on a going concern or at liquidation. The possibility of limiting Tier 1 capital to common equity is now being considered;

3. Build up buffers in good times that can be drawn on in periods of stress: Reducing the procyclicality of bank behaviour is vital. Risk weights calculated on short series that do not include downturns are thus of little value. Banks have to provision for potential losses through the cycle;

4. Regulation of systemically important banks more stringent than for other banks: This can be achieved by introducing overall higher capital requirements or seeking to restrict growth in these institutions by imposing stricter requirements than for other banks;

5. Increased and improved liquidity: The financial crisis showed that many assets held by banks for liquidity purposes were not particularly liquid. More capital is important for the solidity of the banking sector. The Basel rules require capital for banks' assets, but do not stipulate how the assets are to be financed or their liquidity. It is now being considered whether common requirements should be introduced as to the amount of liquid assets a bank must hold to survive a stress situation over a period of for example 30 days without drawing on loans in the central bank. In addition, there will be funding stability requirements. The Basel Committee is expected to make concrete proposals in the course of the year.

Glenn Stevens, Governor of the Reserve Bank of Australia discusses the perspectives on the financial regulation in Stevens (2009): I suggest that we need to do two things.

1. First, we need to draw the right lessons from our experience of the past couple of years. There are no doubt many lessons that might be mentioned. I will list just a few that I think are important:

1) First lesson is that the business cycle still exists and that financial behaviour matters, sometimes a lot, to how that cycle unfolds. We need to have a broad definition of the term "business cycle" in mind. There is more than just the cycle in the "real" economy of GDP, employment, consumer price inflation and so on. These remain important, but it is just as important to recognize the cycles in risk-taking behaviour and finance. Failing to do that is precisely what has got some countries into trouble this time;

2) Second lesson we ought to draw from recent experience is that while downturns will inevitably occur, we are not helpless to do anything about their severity;

3) Next lesson: when downturns come, we can recover;

4) Long-term investments in prudent fiscal and monetary frameworks paid off.

2. Second, we need to apply those lessons in the right way to the challenges that are likely to confront us over the years ahead:

1) First, we start this upswing with less spare capacity than some previous ones. After a big recession, it usually takes some years for well-above-trend growth in demand to use up the spare capacity created by the recession. This time that process will not take as long. Most measures of capacity utilization, unemployment and underemployment are much more like what we saw after the slowdown in 2001, than what we saw after the recession in the early 1990s;

2) Second, and following on the theme of potential supply, others have noted that the rate of population growth at present is the highest since the 1960s. On one hand, this may help alleviate capacity constraints, insofar as certain types of labor are concerned. On the other hand, immigrants need to house themselves and need access to various goods and services as well. That is, they add to demand as well as to supply;

3) Third, the likely build-up in resources sector investment over the years ahead carries significant implications for the medium-term performance and structure of the economy;

4) Australia has many advantages. The financial sector remains in pretty good shape. The Government does not own, and has not had to give direct support to, any financial institution. Australia, therefore, will be relatively free of the difficult governance and exit strategy challenges that such support is raising in some countries. Public finances remain in good shape, with a medium-term path for the budget back towards balance, and without the large debt burdens that will inevitably narrow the options available to governments in other countries. We remain open for trade and investment, with an exposure to Asia, which still has the most dynamic growth potential in the world over the next several decades.

Mark Carney, Governor of the Bank of Canada makes his thoughtful remarks on the problem in Carney (2009): There are two main approaches to reform:

1. First, protect the banks from the economic cycle; in other words, make each bank, individually, more resilient:

i. Raise the quality, consistency, and transparency of the Tier 1 capital base. Going forward, the predominant form of Tier 1 capital must consist of common shares and retained earnings. Moreover, deductions and prudential filters (such as goodwill and other intangibles, investments in own shares, deferred tax assets, etc.) will be harmonized internationally and generally applied at the level of common equity. Finally, all components of the capital base will be fully disclosed.

ii. Introduce a leverage ratio as a supplementary measure of capital adequacy to the Basel II risk-based framework. To ensure comparability, the details of the leverage ratio will be harmonized internationally, fully adjusting for differences in accounting (such as netting).

iii. Introduce a framework for countercyclical capital buffers above the minimum requirement. The Bank of Canada is working closely with OSFI and our international counterparts on proposed elements of this framework.

iv. Create a minimum global standard for funding liquidity that includes requirement for a stressed liquidity-coverage ratio, underpinned by a longer-term structural liquidity ratio. As in other areas, standard setters will need to take a comprehensive approach.

Central banks have real concerns about ratchet effects (large buffers are built beyond the standard), particularly in light of the inherent procyclicality of liquidity (i.e., institutions want more liquidity in bad times so "buffers" cannot be drawn upon).

One potential mitigant would be to ensure that there is a broad range of securities that are liquid in all states of the world. That is tougher than it sounds as anyone who tried to repo quasi-sovereigns throughout the last year knows. I will address some potential solutions to this problem momentarily. The Bank strongly believes that the standard should not bind in times of systemic crises.

2. Second, protect the cycle from the banks; that is, make the system as a whole more resilient:

i. As is the case in Canada, all regulators should institute staged intervention regimes to detect problems early.

ii. Banks themselves should develop "living wills," or plans to unwind in an orderly fashion if they were to fail. If this process results in simpler organizations, so be it. At a minimum, the exercise will underscore the shared responsibility for financial stability and improve regulators' understanding of firms' business models.

iii. The Basel oversight committee agreed to "reduce the systemic risk associated with the resolution of cross-border banks." Closing down a multinational institution is a horrifically difficult challenge, but without progress in this area, the efficiency of the global system will likely decline, perhaps significantly. For example, viable cross-border resolution is the key to ensure that a financial institution's liquidity continues to be optimally distributed across its international operations.

iv. Finally, the Bank of Canada believes that continuously open markets are essential for a system to be robust to failure. The crisis was clearly exacerbated by the seizure of interbank and repo markets. Good collateral became un-financeable overnight, firms failed, risk aversion skyrocketed, and the global economy plummeted.

Stefan Ingves, Governor of the Sveriges Riksbank elaborates on monetary policy and financial stability in Ingves (2009):

1. Monetary policy and the economic situation: the policy we are currently conducting is what we consider to be necessary to attain the inflation target and to provide sufficient support for the recovery in the Swedish economy. We cannot adapt monetary policy on the basis of individual markets as long as we do not regard this as something that affects our ability to meet the inflation target and to safeguard financial stability;

2. "Extraordinary" or "unconventional" measures to attain the inflation target and to safeguard financial stability: Governments, central banks and other authorities around the world have taken extensive measures to recreate confidence in the financial markets, to maintain financial stability and to counteract the negative effects of the crisis on economic activity. The central banks' measures, such as lending to the banks, are reflected in the fact that their balance sheets have increased substantially during the crisis:

1) To attain the inflation target, it has been necessary to make use of new tools to get the financial markets functioning better, to increase the access to loans and to try to bring down various interest rates and risk premiums that have counteracted monetary policy. One example is that we have clearly communicated that we assume that the repo rate will remain at 0.25 per cent for a fairly long period of time. In our most recent forecast the repo rate remains around this level until autumn 2010. Another measure is that the Riksbank lends money to the banks at longer maturities than normal. The purpose is to contribute to continued lower interest rates on loans to companies and households;

2) The measures implemented to safeguard financial stability have similar effects, but the fundamental aim is different. The intention has mainly been to make it easier for the Swedish banks to manage their short-term and medium-term funding. Particularly during the most intense phase of the crisis it was difficult for the banks to borrow other than at very short maturities. It may also be necessary to help financial markets that have suffered serious disruptions to function more smoothly. One can say that the Riksbank has functioned as an intermediary in the bank system – a function that the market does not normally require. One BIS Review 138/2009  7  example in this area, too, is the loans offered to the banks at longer maturities, in both Swedish kronor and US dollars. Another example is that the Riksbank has entered into what are known as swap agreements with the Fed and the ECB, which means that if necessary we can borrow dollars or euro in exchange for Swedish kronor. A further measure is that we accept more types of security as collateral when the banks borrow money from the Riksbank. Moreover, the Riksbank has in different contexts extended the list of eligible counterparties so that more financial agents will have the chance of easier management of liquidity and funding;

3. Change is required in a number of different areas. The most recent crisis has demonstrated how a liquid market can very quickly dry up and become illiquid under certain conditions. Tougher rules are required regarding liquidity and the buffer that banks and credit institutions need to be able to withstand shocks. Further consideration is required here before we find a good solution. It is also necessary to develop further the analysis of asset prices and how the financial markets can be better integrated into our models and forecasts. This applies both with regard to the work on financial stability and the work on monetary policy.

We have completed the review on initiatives toward the financial regulation reform by central bankers. So, we can focus on our original innovative propositions now.

Let us say that the attempts to understand the nature of business cycles in finances and economics with the application of statistical models for money, wealth, and income distributions developed in the econophysics - a new interdisciplinary research field applying methods of statistical physics to problems in economics and finance - were made in a number of researches in Stanley (1996); Yakovenko, Rosser (2009). In the econophysics, the researchers used the knowledge of physics to get an insight on the problems frequently observed in economics in Mosekilde (1996). However, the econophysicists didn't attempt to propose the remedies to stabilize the highly nonlinear characteristics of international financial system behaviour during business cycles.

We propose the Random Tax and Quantum Tax to be considered by G20 financial regulation and taxation experts for possible introduction in taxation regulation of financial institutions with the purpose to encourage banks and hedge funds to pursue the long term investment strategies in productive capacities. We believe that the Random Tax and Quantum Tax are the remedies, which will solve most of the problems in international financial system and change the competitive landscape in banking to the best.

We propose that:

**1. The Random Tax must be imposed on the profits obtained by investment banks as a result of their random decisions to make the high-risk high-profit speculative investment transactions in equity markets, which tend to increase the volatility, originate the bubbles and destabilize the free market itself.**

**2. The Quantum Tax has to be imposed on the profits obtained by hedge funds, which implement the investment strategies toward the search for first signs of investment bubbles creation in equity markets, and then, intentionally invest the big capitals in discovered investment bubbles with the aim to get huge**

**speculative profits resulting in free market destabilization and possible collapse. The basic idea of Quantum Tax is to take the quanta of profit made by every hedge fund, which had hugely benefited from the investment bubbles creation and development in the equity or money markets, back to the state.**

The original proposition on the Random Tax was discussed by authors with Martin Wolf recently, who called it the Windfall Tax and supported the idea of the Random Tax / Windfall Tax introduction in Wolf (2009):

1. All the institutions making exceptional profits do so because they are beneficiaries of unlimited state insurance for themselves and their counterparties;

2. The profits being made today are in large part the fruit of the free money provided by the central bank, an arm of the state;

3. The case for generous subventions is to restore the financial system – and so the economy – to health. It is not to enrich bankers, particularly not those engaged in the sorts of trading activities that destroyed the financial system in the first place;

4. Ordinary people can accept that risk takers receive huge rewards. But such rewards for those who have been rescued by the state and bear substantial responsibility for the crisis are surely intolerable. Bankers are about to reap huge rewards. This damages the legitimacy of the market economy;

5. It is hard to argue in favor of exceptional interventions to bail out the financial sector at times of crisis, and also against exceptional interventions to recoup costs when the crisis is past. "Windfall" support should be matched by windfall taxes;

6. These are genuine windfalls. They are, as George Soros has said, "hidden gifts" from the state. What the state gives, the state is entitled to take back, if it is not used for the state's purposes.

We conclude with a remark that the global financial crisis has already changed the competitive landscape in banking and facilitated the introduction of meaningful banking regulation reform aiming the creation of evolutive competitive environment to maximize the effectiveness of international financial system through the adaptation of fair competition process among the banks in free market capitalism. Going from the 1888 FT's motto: "Without fear and without favour," we came up with the innovative initiative to propose that the Random Tax and the Quantum Tax to be considered by the G20 financial regulation and taxation experts for possible introduction in taxation regulation of financial institutions with the purpose to encourage banks and hedge funds to pursue the long term investment strategies in productive capacities and discourage the gambling in finances. We are confident that the International Centre for Financial Regulation and Financial Times will evaluate the usefulness of the Random Tax and Quantum Tax introduction during the banking and tax regulation reforms and support our innovative initiative to improve the competitive landscape in banking in XXI century.